\documentclass[twocolumn,aps,prX,superscriptaddress]{revtex4-1}
\usepackage{graphicx}
\usepackage{amsmath}
\usepackage{amssymb}
\usepackage{hyperref}
\usepackage{color}
\usepackage{ulem}

\newcommand{\beq}{\begin{equation}}
\newcommand{\eeq}{\end{equation}}
\newcommand{\beqn}{\begin{eqnarray}}
\newcommand{\eeqn}{\end{eqnarray}}
\graphicspath{{Figures/}}

\begin{document}
\title{Unfolding multi-particle quantum correlations hidden in decoherence}
\author{Ke-Ji Chen}
\thanks{They contribute equally to this work. }
\affiliation{Department of Physics, The Chinese University of Hong Kong, Shatin, New Territories, Hong Kong SAR, China}
\affiliation{Department of Physics, Tsinghua University, Beijing,  100084, China}
\author{Ho Kwan Lau}
\thanks{They contribute equally to this work. }
\affiliation{Department of Physics, The Chinese University of Hong Kong, Shatin, New Territories, Hong Kong SAR, China}
\author{Hon Ming Chan}
\affiliation{Department of Physics, The Chinese University of Hong Kong, Shatin, New Territories, Hong Kong SAR, China}
\author{ Dajun Wang}
\affiliation{Department of Physics, The Chinese University of Hong Kong, Shatin, New Territories, Hong Kong SAR, China}
\author{Qi Zhou}
\email{zhou753@purdue.edu}
\affiliation{Department of Physics and Astronomy, Purdue University, West Lafayette, IN, 47906, US}
\date{\today}
\begin{abstract}

Quantum coherence is a fundamental  characteristic to distinguish quantum systems from their classical counterparts. Though quantum coherence persists in isolated non-interacting systems, interactions inevitably lead to decoherence, which is in general believed to cause the lost of quantum correlations.  Here, we show that, accompanying to the single-particle decoherence, interactions build up quantum correlations on the two-, three-, and multi-particle levels. Using the quantitative solutions of the quantum dynamics of a condensate occupying two modes, such as two bands of an optical lattice, we find out that such dynamically emergent multi-particle correlations not only reveal how interactions control the quantum coherence of a many-body system in a highly intriguing means, but also evince the rise of exotic fragmented condensates, which are difficult to access at the ground state. We further develop a generic interferometry that can be used in experiments to measure high order correlation functions directly. 

\end{abstract}

\maketitle

\section{introduction}
Because of the superposition principle,  quantum coherence of an isolated single particle naturally persists forever. For instance, an isolated single spin processes in a magnetic field, and  the spin coherence, which is characterised by the transverse magnetisation, never decays \cite{slichter-90}. In a many-body system, phenomena associated with quantum coherence become much richer \cite{ streltsov-16,  Bloch-08}. Whereas the well developed spin-echo techniques overcome the dephasing due to inhomogeneous external fields \cite{slichter-90, hahn-50, purcell-54}, introducing interactions to the problem makes it highly nontrivial \cite{sarma-03, demler-08, ma-14, wei-12, peng-14, yang-17}. When a particle interacts with either the environment, or other particles in the same quantum system, even sophisticated extensions of spin echo techniques could only partially restore the quantum coherence \cite{demler-08, yan-13, yao-07}. It is in general believed that interactions lead to unavoidable decoherence such that quantum coherence and correlations are lost \cite{leggett-87, stamp-00, zurek-03}.

Ultracold atoms provide physicists an ideal platform to explore quantum many-body dynamics, due to its weak coupling to environment and the absence of disorders \cite{bloch-08}. An ultracold atomic cloud can be essentially regarded as an isolated system, and allows physicists to trace a wide range of non-equilibrium phenomena \cite{Ho-06, eisert-15, altman-15, polkovnikov-11}.  In particular, a Bose-Einstein condensate (BEC) allows one  to study the quantum dynamics at a macroscopic level. Quantum coherent dynamics has been observed in a variety of systems~\cite{chapman-05, ketterle-97, cornell-98,  bloch-02, chapman-12, cheng-13, wang-15, cheng-17}. However,   interaction induced quantum decoherence remains a challenge for both theorists and experimentalists, as it is notably difficult to trace the many-body quantum dynamics. A fundamental question naturally arises, what is the fate of  quantum correlations after the quantum decoherence occurs in such isolated interacting quantum many-body systems?

\begin{figure}
\begin{center}
\includegraphics[width=3.3in]{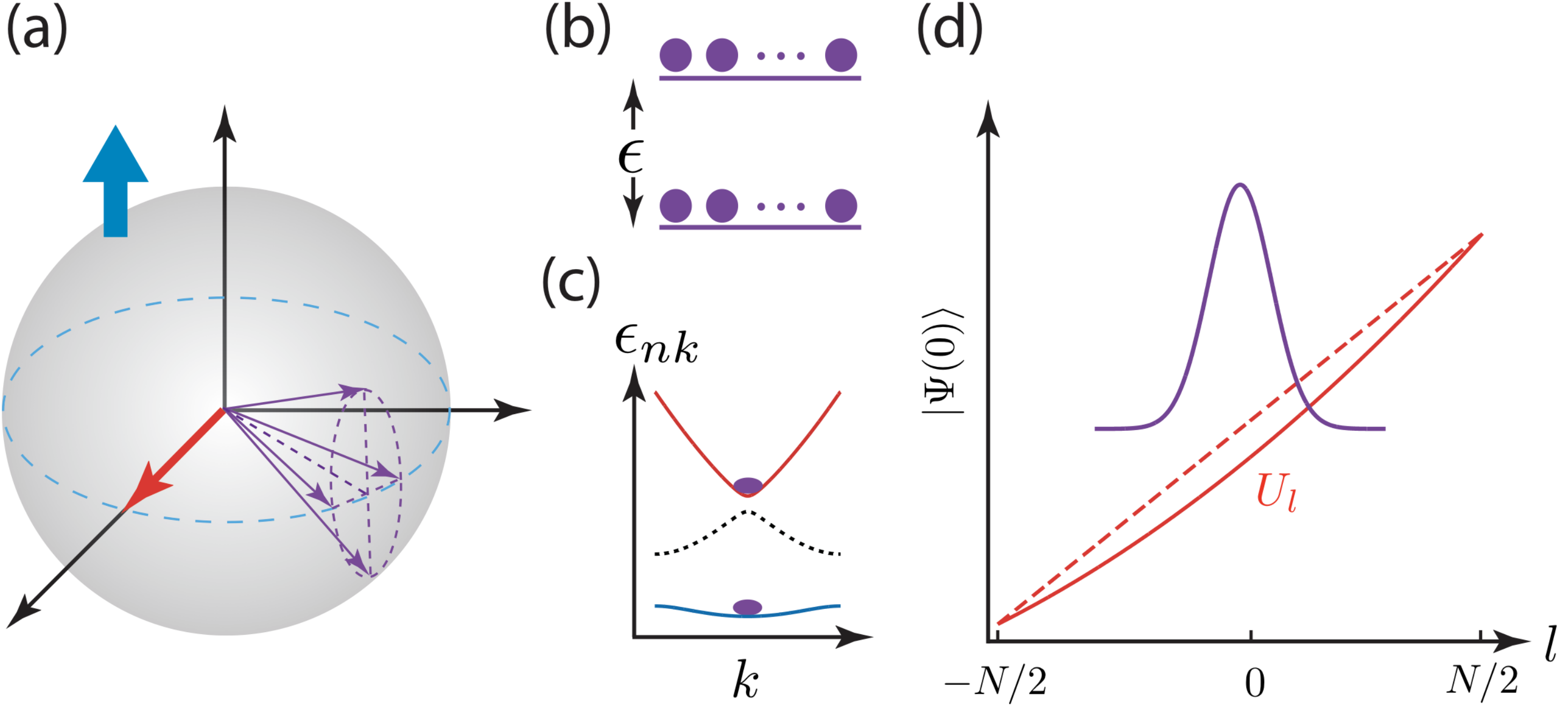}
\caption{(a)  At the initial time, a collection of spins (thick red arrow)  points to the same direction on the Bloch sphere. While each spin (thin purple arrow) processes about the $z$ axis,  interactions scatter spins to different locations on the Bloch sphere and lead to decoherence.  (b)  The energy splitting $\epsilon$ in a two-level system serves  as an effective magnetic field along the z direction. If a boson (purple sphere) is in state $|1\rangle$ ($|2\rangle)$, it corresponds to spin-up (spin-down). (c) An example of the realization of a two-level system. Bosons are prepared at an initial state that is a superposition of zero crystal momentum states of the $s$ and $d$ band in a one-dimensional optical lattice. (d) The many-body dynamics can be described by an effective one-dimensional model. The time evolution of a wave packet (purple curve) of a fictional particle captures the quantum decoherence and revival of the many-body bosonic state. Dashed line represents the external potential felt by this fictional particle when interactions are absent in the original model. Finite interactions add quadratic corrections  to the external potential (red solid curve). }
\label{Fig1}
\end{center}
\end{figure}

Here, we consider the quantum dynamics of a generic many-body bosonic system, in which $N$ bosons occupy two modes. The initial state is a coherent state $\frac{1}{\sqrt{N!}}(\frac{a_1^\dagger +a_2^\dagger}{\sqrt{2}})^N|0\rangle$, where $a_{i=1,2}^\dagger$ are the creation operators for these two modes.  This initial state can be mapped to $N$ identical psuedospin-1/2s, and the spin coherence is well characterised by the transverse magnetisation, $\sigma_\bot=|\langle a_1^\dagger a_2\rangle |\sim N$, as shown in {Fig.\ref{Fig1}(a)}.  In the absence of interactions, these spins process under an effective magnetic field, which is given by the single-particle energy difference $\epsilon=\epsilon_1-\epsilon_2$ between this two modes, as shown in Fig.\ref{Fig1}(b),  and $\sigma_\bot$ never decays. When there are interactions between these two modes, quantum coherence is indeed suppressed on the single-particle level.  As expected, the single-particle correlation function $\langle a^\dagger_1 a_2\rangle$ decays. However, multi-particle correlations naturally establish themselves as time goes on. We will quantitatively show that at certain times,  the high order correlation functions $\langle a^{\dagger m}_1 a_2^m\rangle$ ($m>1$) become the order of $N^m$, while $\langle a^{\dagger }_1 a_2\rangle$ is suppressed down to zero. This clearly demonstrates the intriguing role of interactions in quantum many-body dynamics, which act as  the source for both the single-particle decoherence and multi-particle correlations. In particular, when multi-particle correlations arise in the absence of single-particle coherence, fragmented condensates emerge. Our work thus sets up a new routine for accessing this type of exotic phases,  which are difficult to access at the ground state.

\section{model and theoretical results}
Whereas our results are quite general, to concretise the discussions, we use two bands in an optical lattice as the two modes to illuminate the physical picture.  Recent experiments have successfully prepared bosons that coherently occupy both the $s$ and $d$ bands \cite{zhou-13, zhou-16}, as shown in Fig.\ref{Fig1}(c).  The condensate wavefunction is written as 
\begin{equation}
|\Psi (0)\rangle=\frac{1}{\sqrt{N!}}\left(\frac{a_{s}^\dagger +a_{d}^\dagger}{\sqrt{2}}\right)^N|0\rangle, \label{is}
\end{equation}
where $a_{s}^\dagger $ ($a_{d}^\dagger $) is the creation operator at the $s$ ($d$) band with zero momentum. Here, we consider weakly interacting bosons and ignore the small condensate depletion at finite momenta, which does not affect the main results in the time scale that is relevant to our discussions. The index for the momentum is thus supressed. In the non-interacting limit, the band gap $E_g$ acts as an effective Zeeman splitting for $N$ identical pseudospin-1/2s, which process with a period $T_0=h/E_g$.  Compared with other two-mode or spin-1/2 systems, the advantage of this system is that,  $E_g$ is much larger than other energy scales in the system.  For rubidium atoms, a lattice depth of $15-20E_R$ has a band gap $E_g$ that is $34-42   h\times 10^3  Hz$. This corresponds to a time scale $T_0$ of the order of a few tens of $\mu s$ \cite{zhou-13, zhou-16}. Compared with other many-body dynamics in ultracold atoms, such as spinor condensates with a typical spin oscillation period of the order of a few hundred $ms$ \cite{spindy}, $T_0$ here is well separated from other time scales, such as the life time of a BEC of the order of $s$ \cite{leggett-01}. This time scale separation allows physicists to access the decoherence purely induced by mutual interactions between atoms, as the effects of the coupling coupling to environment only occur in a much larger time scale.

Without interactions, the $N$ identical pseudospin-1/2s process without decay. Taking into account interactions, the Hamiltonian can be written as
\begin{equation}
\begin{split}
\hat{H}&=E_g \hat{N}_d+g_s \hat{N}^2_s+ g_d \hat{N}^2_d+4g_{sd}\hat{N}_s\hat{N}_d\\
& +(u\hat{N}_s \hat{a}_s^\dagger \hat{a}_d +u' \hat{N}_d\hat{a}_s^\dagger \hat{a}_d+g_{sd} \hat{a}_s^{\dagger 2} \hat{a}_d^{2}+h.c),  \label{H}
\end{split}
\end{equation}
where $\hat{N}_{i=s,d}$ is the number operators for the $s$ or $d$ band, $g_s$ and $g_d$ are the intra-band interactions, and $g_{sd}$ is the inter-band interaction. The last three terms in Eq.(\ref{H}) describe interaction induced density assisted tunnelling and pair tunnelling. Details of how to determine  all coefficients in Eq.(\ref{H}) from the microscopic Hamiltonian are given in the  {appendix \ref{sec:A}}. Eq.(\ref{H})  has included the most general interactions for a two-mode or two-level system. The simple picture of $N$ identical pseudospin-1/2s no longer applies for interacting systems. To evaluate the wavefunction at time $t$, $|\Psi(t)\rangle=e^{-i\frac{\hat{H}}{\hbar}t}|\Psi (0)\rangle$, we expand the initial state in the basis of eigenstates of the Hamiltonian, $|\Psi(0)\rangle=\sum_n \alpha_n(0)|\psi_n\rangle$, where $|\psi_n\rangle$ satisfies $\hat{H}|\psi_n\rangle=E_n|\psi_n\rangle$. Each energy eigenstate  $|\psi_n\rangle$ is written as 
$|\psi\rangle_n=\sum_{l=-N/2}^{N/2} c_{nl} |l\rangle$, where $|l\rangle=\frac{a_s^{\dagger \frac{N}{2}+l} a_d^{\dagger \frac{N}{2}-l}}{\sqrt{(\frac{N}{2}+l)!(\frac{N}{2}-l)!}}  |0\rangle$ is the Fock state. To simplify the notations, we have assumed that $N$ is even, as an even or odd $N$ essentially makes  no difference at large $N$ limit. The matrix representation of $\hat{H}$ is written as 
\begin{equation}
\begin{split}
(E_{nl}-E)|l\rangle=U_l |l\rangle +\sum_{s=1,2} J_{l,l\pm s}|l\pm s\rangle , \label{M}
\end{split}
\end{equation}
where 
\begin{equation}
\begin{split}
&U_l=A_1 l+A_2l^2, \,\,\ A_1=(g_{s}-g_{d})N-E_{g}\approx -E_g,\\
&A_2=g_{s}+g_{d}-4g_{sd},\\
&J_{l,l+1}=\left(u(\frac{N}{2}+l)+u'(\frac{N}{2}-l)\right)\sqrt{(\frac{N}{2}+l+1)(\frac{N}{2}-l)},\\
&J_{l,l+2}=g_{sd}\sqrt{(\frac{N}{2}+l+2)(\frac{N}{2}+l+1)(\frac{N}{2}-l)(\frac{N}{2}-l-1)},
\end{split}
\end{equation}
Eq.(\ref{M}) maps the quantum many-body dynamics to a simple one-dimensional lattice model, as Fig.\ref{Fig1}(d) shows, in which $U_l$ is the onsite energy, $J_{l,l\pm 1}$ and $J_{l,l\pm 2}$ are the nearest and next nearest neighbour tunnelings, respectively. 

\begin{figure}
\begin{center}
\includegraphics[width=3.5in]{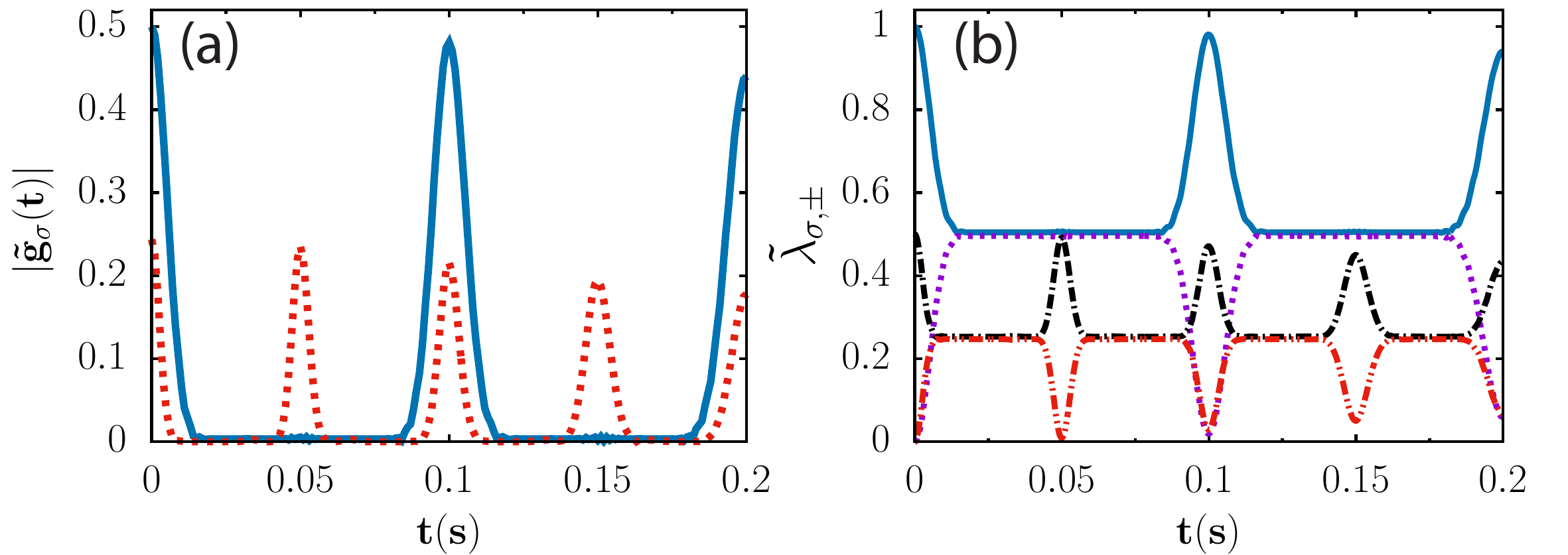}
\caption{(a) One-particle (solid blue) and two-particle (dashed red) correlation functions.  $\tilde{g}_1(t)=g_1(t)/N$ and $\tilde{g}_2(t)=g_2(t)/N^2$. 
(b) Eigenvalues of reduced density matrices $\rho_{1}(t)$ and $\rho_2(t)$. $\tilde{\lambda}_{1,\pm}=\lambda_{1,\pm}/N$ (solid blue and dashed purple) and  $\tilde{\lambda}_{2, \pm}=\lambda_{2, \pm}/N^2$ (black dash-dotted and red dash-double-dotted )  are the renormalized eigenvalues of $\rho_1(t)$ and $\rho_{2}(t)$), respectively.  Parameters are  $N_{\text{tot}}=10^5,V_{\|}=10E_{R}, L_{\|}=L_{\bot}=50, N=40$.  $a_{s}=5.1nm, d=426nm$ and $V_{\bot}=20E_{R}$ is the lattice depth in the $y-z$ plane.} 
\label{Fig2}
\end{center}
\end{figure}

\subsection{Decoherence, revival, and emergent multi-particle correlations and fragmentation}
We solve Eq.(\ref{M}) numerically. Using realistic experimental parameters, the time dependent one-particle correlation function, $g_1(t)\equiv \langle \Psi(t) |b^\dagger_s b_d|\Psi(t)\rangle$, and two-particle correlation function, $g_2(t)\equiv \langle \Psi(t) |b^\dagger_sb^\dagger_s b_db_d|\Psi(t)\rangle$, have been computed and are shown in  {Fig.\ref{Fig2}(a)}, where $|\Psi(t)\rangle$ is the wavefunction at time $t$. A number of characteristic features are clear in this figure. With increasing $t$ from zero,  both $g_1(t)$ and $g_2(t)$ decay fast, in a time scale $\tau_{i=1,2}$, as expected from quantum decoherence in an interacting system. However, these correlation functions revive in a revival time scale $T_{i=1,2}$. The most striking result is that, $g_2(t)$ comes back much earlier than $g_1(t)$, i.e., $T_2=T_1/2$.  At $t=T_2$, the vanishing single-particle correlation $g_1(t)$ and a macroscopic $g_{2}(t)$  signify the rise of an exotic fragmented condensate. When $g_1(t)$ vanishes, the reduced one-particle density matrix, $ \langle \Psi(t) |b^\dagger_\mu b_\nu|\Psi(t)\rangle$, becomes 
\begin{equation}
\rho_1(T_2)=\left(\begin{array}{cc} \langle \hat{N}_{s}(T_2)\rangle & 0 \\0 & \langle \hat{N}_{d}(T_2)\rangle\end{array}\right),
\end{equation}
where {$\langle \hat{N}_{\mu=s,d}(t)\rangle=\langle \Psi(t)| b^\dagger_\mu b_\mu|\Psi(t)\rangle $}. Since $\rho_1(T_2)$ has two eigenvalues of the order of $N$, $|\Psi(T_2)\rangle$ corresponds to a fragmented condensate \cite{penrose-51, penrose-56, James-82, mueller-06, Kang-15}. This is very different from the initial state, $|\Psi(0)\rangle$, whose reduced one-particle density matrix has only one eigenvalue of the order of $N$. Moreover, one could evaluate the reduced two-particle density matrix, which is written as
\begin{equation}
\rho_2(T_2)=\left(\begin{array}{cc}  \langle \hat{N}^2_{s}(T_2)\rangle & g_2(T_2) \\g^{\ast}_2(T_2) &  \langle \hat{N}^2_{d}(T_2)\rangle\end{array}\right),
\end{equation}
where $\langle \hat{N}^2_{\mu=s,d}(T_2)\rangle= \langle \Psi(t)| b^{\dagger 2}_\mu b^2_\mu|\Psi(t)\rangle$. Since all matrix elements of $\rho_2(T_2)$ are of the order of $N^2$, it has only one eigenvalue that is of the order of $N^2$. Whereas $g_1(t)$ characterises the single-particle coherence of the initial state $|\Psi(0)\rangle$, $g_2(t)\sim N^2$ characterises the coherence between two particles in the fragmented condensate at $t=T_2$. Thus, after decoherence occurs, the quantum many-body dynamics produces an exotic state, $|\Psi(T_2)\rangle$, which can be viewed as a pair condensate distinct from the initial single-particle condensate. Fig.\ref{Fig2}(b)  shows the eigenvalues of $\rho_1(t)$ and $\rho_2(t)$ as functions of $t$.   When $t=n T_2$, where $n$ is an odd  integer,  two eigenvalues of $\rho_1(t)$ are both of order $N$. This signify the rise of fragmented condensate.

\subsection{ Constructive and destructive interferences }
To understand the above results, an approach in the zero tunnelling limit is very useful to highlight qualitatively the underlying physics. When $J_{l,l\pm s}=0$, analytical solutions are available and shed light on the underlying mechanism of the coherence and decoherence in the quantum many-body dynamics. Quantitatively, such approach also captures the essentially physics at small times $t$ before finite tunnelings $J_{l,l\pm s}$ affect the results. Apparently, when $J_{l,l\pm s}=0$, Fock states $|l\rangle$ become the eigenstates of $\hat{H}$, i.e., $c_{nl}=\delta_{nl}$, and the eigenenergy is simply the onsite energy $E_n= U_l\delta_{nl}$. The expansion of the initial state can be written as  $|\Psi(0)\rangle=\sum_l \alpha_l(0)|l\rangle$. From Eq.(\ref{is}), it is clear that such expansion corresponds to a binomial distribution, {$\alpha_l(0)=\frac{1}{\sqrt{2^NN!}}C_N^{\frac{N}{2}-l}\sqrt{(\frac{N}{2}+l)!(\frac{N}{2}-l)!}=\left({C_N^{\frac{N}{2}-l}}/{2^N}\right)^{\frac{1}{2}}$}. The wavefunction at time $t$ is written as 
\begin{equation}
|\Psi(t)\rangle=\sum_{l=-\frac{N}{2}}^{\frac{N}{2}}\alpha(t) |l\rangle,\,\,\,\,\,\,\, \alpha(t)=e^{-i\frac{U_l t}{\hbar} } \left(\frac{C_N^{\frac{N}{2}-l}}{2^N}\right)^{\frac{1}{2}}.\label{psit}
\end{equation}
As time goes {on}, interactions give rise to a different dynamical phase factor to each Fock state. These dynamical phase factors control the correlation functions of the many-body system. To characterise the coherence, we evaluate the one-particle correlation function 
\begin{equation}
g_1(t)\equiv \langle \Psi(t) |b^\dagger_s b_d|\Psi(t)\rangle=\sum_l \alpha_{l+1}^*(t)\alpha_l(t)\mathcal{W}_l, \label{sg1}
\end{equation}
where  $\mathcal{W}_l=\sqrt{(\frac{N}{2}+l+1)(\frac{N}{2}-l)}$.  Similarly, multi-particle correlation functions can also be evaluated. For instance, the two-particle correlation function, 
\begin{equation}
g_2(t)\equiv \langle \Psi(t) |b^\dagger_sb^\dagger_s b_db_d|\Psi(t)\rangle=\sum_l \alpha_{l+2}^*(t)\alpha_{l}(t)\mathcal{V}_l,  \label{sg2}
\end{equation} 
where $\mathcal{V}_l=\sqrt{(\frac{N}{2}+l+2)(\frac{N}{2}+l+1)(\frac{N}{2}-l)(\frac{N}{2}-l-1)}$. 

Both $g_1(t)$ and $g_2(t)$ can be expressed in much more illuminating means. In large $N$ limit, a binomial distribution can be well approximated by  a Gaussian, {$\frac{1}{2^N}C_N^k\approx\sqrt{\frac{2}{\pi N}}e^{-\frac{2}{N}(k-\frac{N}{2})^2}$}.  Meanwhile, using the Possion summation formula, we obtain an identity {$\sum^{\infty}_{n=-\infty}e^{-\pi b(n+a)^{2}}=\sum^{\infty}_{n=-\infty}\frac{1}{\sqrt{b}}e^{-\frac{\pi n^{2}}{b}}e^{2\pi i n a}
$}.  Using these two expressions, it is straightforward to rewrite $g_1(t)$ as
\begin{equation}
g_{1}(t)=\frac{N}{2}\sum^{\infty}_{n=-\infty}e^{-\frac{N-1}{2}(\frac{A_{2}t}{\hbar}-n\pi)^{2}}e^{i n\pi}e^{i\frac{A_{1}t}{\hbar}}. \label{g1}
\end{equation}
For non-interacting systems, $A_2=0$. It is rather clear that $|g_1(t)|$ is a time-independent constant. As the $N$ identical pseudospin-1/2 process at the same frequency $\sim 1/E_g$, the transverse magnetisation never decays. When interactions are present, $A_2$ becomes finite, and $g_1(t)$ is a sum of an infinite number of equally spaced Gaussian packets in the time domain. The width of each Gaussian packet and the separation between two nearest packets correspond to two characteristic time scales,  
\begin{equation}
\tau_{1}=\frac{\hbar}{\sqrt{N-1}A_2}, \,\,\,\,\,\, T_{1}=\frac{\pi \hbar}{A_2},
\end{equation}
where the subscript $1$ denotes that a time scale of $g_1(t)$. $\tau_{1}$ is precisely the decoherence time of the one-particle correlation. For a system with a large particle number $N$, the one-particle coherence, i.e., the transverse magnetisation $\sigma_\bot$ in the spin model, quickly decays in a time scale $\tau_{1}$.  $T_{1}$ sets up another time scale, the revival time, at which $g_1(t)$ recovers the original value $g_1(0)$. 

The result of $g_1(t)$ is consistent with our expectation that interactions inevitably lead to decoherence.  However, the nature of the quantum many-body state in the time domain $\tau_{1} \ll t \ll T_{1}$ remains unclear, unless we continue to explore $g_2(t)$ and even higher order correlation functions.  Here, $g_2(t)$ can be evaluated using the same techniques for $g_1(t)$. We obtain,
\begin{equation}
g_{2}(t)=\frac{N(N-1)}{4}\sum^{\infty}_{n=-\infty}e^{-\frac{N-2}{2}(\frac{2A_{2}t}{\hbar}-n \pi)^{2}}e^{i\frac{2A_{1}t}{\hbar}} .\label{g2}
\end{equation}
Clearly, Eq.(\ref{g2}) has the same structure as Eq.(\ref{g1}). We define the decoherence time and the revival time for $g_2(t)$,
\begin{equation}
\tau_{2}=\frac{\hbar}{2\sqrt{N-2}A_2}, \,\,\,\,\,\, T_{2}=\frac{\pi\hbar}{2A_2},
\end{equation}
where the subscript $2$ denotes a time scale of $g_2(t)$. It is clear that  $T_2$ halves $T_1$. Since $\tau_1 \ll T_2 <T_1$ is well satisfied in large $N$ limit, we reach an important conclusion that, at time $T_2$, the single-particle correlation is suppressed down to zero and two-particle correlation function becomes the order of $N^2$, i.e., $g_1(T_2)=0$ and $g_2(T_2)\sim N^2$. 

\begin{figure}
\begin{center}
\includegraphics[width=3.2in]{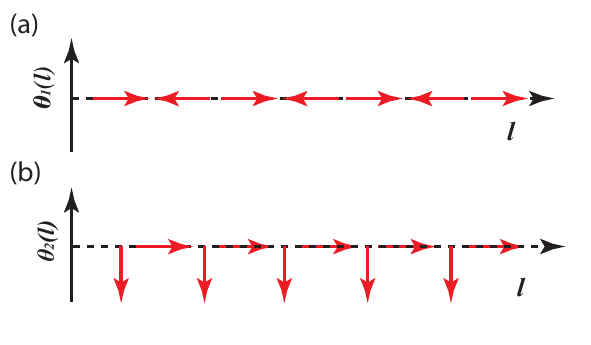}
\caption{Schematic of the phase of the Fock states as a function of $l$, where $\theta_{m}(l)=e^{-i\frac{\pi l^2}{m}}$. (a) $\theta_{1}(l)$ when $t=T_1=\frac{\pi \hbar}{A_{2}}$.  Both one-particle and two-particle correlation functions reach maximal values. (b) $\theta_2(l)$ when $t=T_2=\frac{\pi\hbar}{2A_{2}}$. One-particle correlation function is suppressed down to zero while two-particle correlation function is maximized.}
\label{Fig3}
\end{center}
\end{figure}

$g_1(t)$ and $g_2(t)$ at both $T_1$ and $T_2$ can be qualitatively understood from the following picture. If we use a two-dimensional unit vector to represent the time-depedent phase of each Fock state, each vector rotates under a local effective magnetic field, which is given by $U_l$ defined in the one-dimensional lattice in Eq.(\ref{M}). In the absence of interactions, $U_l$ is linear, $|g_1(t)|$ never vanishes.  However, for finite interactions, $A_2\neq 0$.   At $t=T_1$, from Eq.(\ref{psit}), we see that {$\alpha_l(t)\sim \theta_1(l)e^{-i\frac{A_1}{A_2}\pi l}$,  where $\theta_1(l)=e^{-i\pi l^2}$, since  $e^{-i\frac{A_1}{A_2}\pi l}$ does not affect the amplitude of correlation functions,  and $\theta_1(l)$ is curial. $\theta_1(l)$ can be rewritten as $e^{-i\pi l}$}, i.e., the phase increases linearly with increasing $l$. When evaluating $g_1(t)$ in Eq.(\ref{sg1}), different terms add constructively. Both $g_1(t)$ and $g_2(t)$ are maximized at $t=T_1$, as shown in {Fig.\ref{Fig3}(a)}. When $t=T_2$, $\alpha_l(t)\sim \theta_2(l)e^{-i\frac{A_1}{2A_2}\pi l}$, where $\theta_2(l)=e^{-i\pi l^2 /2}$. Whereas $\alpha_{l\in even} (t)\sim e^{-i\frac{A_1}{2A_2}\pi l}$, there is an extra phase of $\pi/2$ for odd $l$, $\alpha_{l\in odd}(t)\sim  -i e^{-i\frac{A_1}{2A_2}\pi l}$, as shown in {Fig.\ref{Fig3}(b)}. Thus, the contributions to $g_1(t)$ from the $l$th and $l+1$th term in Eq.(\ref{sg2}) essentially cancel each other due to a destructive interference. In contrast, $g_2(t)$ is not affected as what enters Eq.(\ref{sg2}) is $\alpha_{l+2}^*(t)\alpha_{l}(t)$. Thus, $g_2(t)$ is maximized at $T_2$. 

The above discussions can be directly generalised to $m$-particle correlation functions. We find out that the  decoherence time and the revival time for $g_m(t)\equiv \langle \Psi(t) |b^{\dagger m}_s b^m_d|\Psi(t)\rangle$ can be written as
\begin{equation}
\tau_{m}=\frac{\hbar}{m\sqrt{N-m} A_2}, \,\,\,\,\,\, T_{m}=\frac{\pi \hbar}{mA_2}.
\end{equation}
Thus at time $T_m$, correlation functions $g_{m'}(T_m)$ vanish, if $m' \mod m \neq 0$. In contrast, $g_{m'}(T_m)$ become macroscopic if $m' \mod m = 0$, i.e., $g_{m'}(T_m)\sim N^{m'}$. Similar to $g_2(T_2)$, such correlation functions can be understood easily in the zero tunneling limit (see appendix \ref{sec:B}). These results reflect the intriguing interplay between interactions and quantum coherence in an isolated many-body system. After single-particle decoherence occurs, many-particle correlations are inevitably established by interactions in a quantum many-body dynamical evolution.  Consequently, exotic fragmented states arise, in which reduced $m'$-body reduced matrix has multiple macroscopic eigenvalues. If we review the state $|\Psi(T_2)\rangle$ as a pair condensate, then $|\Psi(T_m)\rangle$ can be regarded as a multi-particle condensate. 

It has been well known fragmented condensates exhibit extraordinary properties absent in ordinary condensates, such as large number fluctuations and squeezed spins \cite{mueller-06, ho-00, ho-04}. However, it is challenging to realise a fragmented condensate at the ground state due to the instability against to small external perturbations \cite{mueller-06}. Here, fragmented condensates are generated in a quantum dynamical envolution. The instability issues at the ground state, or more generally, in equilibrium states, are no longer relevant. Here, to observe fragmented condensates $|\Psi(T_m)\rangle$, $T_m$ should be within the time scales accessible in current experiments. Also, the width of the Gaussian packets in the time domain should be large enough for implementing detections or operations in practise. Using realistic experimental parameters for Rb in a 3D lattice, we have found out that for $m=2$ and $m=3$, $T_m$ ($\tau_m$) could be {47$ms$(2.4$ms$) and 31$ms$ (1.65$ms$)}, respectively. All these numbers are accessible in current experiments. In general, for larger $m$, $\tau_m$ becomes smaller.  On the other hand, one could tune both the scattering length and the lattice depth to control $A_2$ so as to increase $\tau_m$. Thus, this provides physicists a new means to access the long-sought fragmented condensates. 

Whereas the zero tunneling approximation has readily provided us a nice description of the underlying physics, we have also applied a more rigorous analytical calculation for finite tunnelings, which is presented in  appendix \ref{sec:C}. This method also gives a qualitative estimation of the small residue $g_1(t)$ at $t=T_2$.  At large $t$, the suppression of the maxium of the correlation functions, or the imperfections of the revival can also been understood by taking into account high order corrections (appendix \ref{sec:D}). 

\section{Measuring multi-particle correlations}
We now discuss how to measure high order correlation functions. It is known that the relative phase in a wavefunction, which controls $g_1(t)$ and other correlation functions, cannot be directly measured from density or populations in each mode. Nevertheless,  a $\pi/2$ pulse can be used to measure $g_1(t)$, or equivalently the transverse magnetization. For instance, for a coherent state $(\alpha \hat{a}_s^\dagger+\beta \hat{a}_d^\dagger)^N|0\rangle$, where $|\alpha|^2+|\beta|^2=1$, a  $\pi/2$ pulse corresponds to a transformation, $\hat{a}_s^\dagger\rightarrow (\hat{a}_s^\dagger+\hat{a}_d^\dagger)/\sqrt{2}$,  $\hat{a}_d^\dagger\rightarrow (\hat{a}_s^\dagger-\hat{a}_d^\dagger)/\sqrt{2}$, and the state becomes $(\frac{\alpha+\beta}{\sqrt{2}} \hat{a}_s^\dagger+\frac{\alpha-\beta}{\sqrt{2}} \hat{a}_d^\dagger)^N|0\rangle$.  The population difference between the $s$ and $d$ bands, or the magentization along the $z$ direction, of the new state directly tells one the spin coherence $\alpha^*\beta$ of the original state. To measure high order correlation functions, we consider a generalized $\pi/2$ pulse $\mathcal{P}_\theta$, which is defined as 
\begin{equation}
\hat{a}_s^\dagger\rightarrow \frac{1}{\sqrt{2}}(\hat{a}_s^\dagger+\hat{a}_d^\dagger),\,\,\,\,\, e^{i\theta} \hat{a}_d^\dagger\rightarrow \frac{1}{\sqrt{2}}(\hat{a}_s^\dagger+\hat{a}_d^\dagger), \label{gpih}
\end{equation}
where $\theta$ corresponds to a ``delayed" $\pi/2$ pulse in our lattice system. For an arbitrary many-body wavefunction, $|\Psi(t)\rangle=\sum_{mn} \frac{\alpha_{mn}(t)}{\sqrt{m!n!}}a^{\dagger m}_s a_d^{\dagger n}|0\rangle$, after a small time $\delta t\sim h/E_g$, interaction effects, which will occur in a much larger time scale, do not change the wavefunction. The only change is that the single particle wavefunction in the $d$ band acquires an additional dynamical phase $\sim e^{-i \frac{E_g}{\hbar} \delta t}$. This corresponds to a transformation of the operator $a_d^\dagger\rightarrow e^{-i \frac{E_g}{\hbar} \delta t}a_d^{\dagger }$. Thus the many-body wavefunction becomes $|\Psi(t+\delta t)\rangle=\sum_{mn} \frac{\alpha_{mn}(t)}{\sqrt{m!n!}}a^{\dagger m}_s (e^{-i \frac{E_g}{\hbar} \delta t}a_d)^{\dagger n}|0\rangle$. Applying a $\pi/2$ pulse to the state $|\Psi(t+\delta t)\rangle$ then corresponds to a  generalized $\pi/2$ pulse applied to state $|\Psi(t)\rangle$. Whereas in this optical lattice system, $\theta$ in Eq.(\ref{gpih}) can be naturally implemented using the aforementioned ``delayed" scheme, in a generic two-mode or two-level system,  a strong effective magnetic field along the $z$ direction could be introduced before the $\pi/2$ pulse. One of the operators then gains an extra dynamical phase, and the transformation in Eq.(\ref{gpih}) can be realised. 

Here we take $g_2(t)$ as an example. After a pulse, the corresponding transformations of the wavefunction are written as
\begin{equation}
|\Psi(t)\rangle \rightarrow |\Psi'_\theta(t)\rangle.
\end{equation}
The density-density correlation functions of the new state, $|\Psi'(t)\rangle_\theta$, could then be measured. Define 
\begin{equation}
F_\theta=\langle \Psi_\theta'(t)|\left(\hat{N}_s(\hat{N}_s-1)+\hat{N}_d(\hat{N}_d-1)-2 \hat{N}_s\hat{N}_d \right) |\Psi'_\theta(t)\rangle,
\end{equation}
a straightforward calculation shows that
\begin{equation}
\text{Re} g_2(t)=\frac{1}{4}(F_{0}-F_{\frac{\pi}{2}}), \,\,\,\, \text{Im} g_2(t)=\frac{1}{4}\left(2F_{\frac{\pi}{4}}-F_{0}-F_{\frac{\pi}{2}}\right).
\end{equation}
Thus, we see that three repeated experiments, which correspond to three generalized $\pi/2$ pulses, $\mathcal{P}_0$, $\mathcal{P}_{\pi/4}$ and $\mathcal{P}_{\pi/2}$, allow one  get $g_2(t)$.
\section{conclusions}
The study of quantum coherence and decoherence is one of the most fundamental problems in modern physics. Whereas it is well accepted that interactions lead to decoherence, we have shown that there is much richer physics behind the decoherence. Though it may be expected that quantum correlations get lost after decoherence occurs, we find out that, exotic states that are characterised by multi-particle quantum correlations arise. Whereas we have been focusing on a particular realization of our model in optical lattices, our results apply to arbitrary many-body bosonic systems described by this generic two-mode model. We hope that our work will stimulate more works to study intriguing quantum correlations hidden in decoherence. We also hope that our work provides physicists a new means to create exotic quantum phases not accessible at equilibrium using quantum many-body dynamics.

\begin{acknowledgements}
QZ acknowledge Xiaoji Zhou and Dan Stamper-Kurn for discussions. This work is supported by startup funds from Purdue University and RGC/GRF 14304015. Part of the manuscript was completed at the Aspen Center for Physics, which is supported by National Science Foundation grant PHY-1607761. 
\end{acknowledgements}

\appendix
\begin{figure*}[t]
\begin{center}
\includegraphics[width=6.9in]{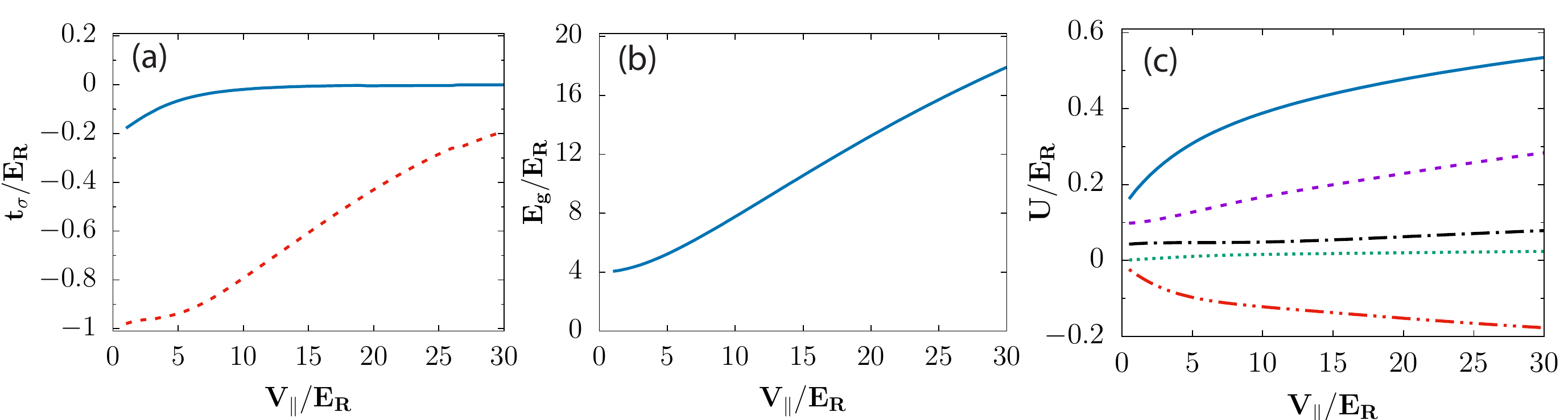}
\caption{(a):  tunnelings of $s$-band (solid bule) and  $d$-band (dashed red) in the $x$ direction  when $V_{\|}$  changes.  (b):  the band gap between  $s$ and $d$ bands. (c): Interaction strengths of 3D optical lattice when $V_{\|}$ changes, $V_{\bot}=20E_{\bf R}$ is fixed, solid bule (dashed purple) curve denotes  intra-band interaction $U_s(U_d)$, double-dotted red (dotted green) and dotted black curves represent $V_a (V_b)$  and $U_{sd}$, respectively. Other parameters are all the same,  i.e. $a_{s}=5.1nm, d=426nm$ and $m=1.443\times 10^{-25}kg$.}
\label{s1}
\end{center}
\end{figure*}

\section{effective hamiltonian}
\label{sec:A} 
$\hat{H}$ in Eq.(2) of the main text is the Hamiltonian describing a generic two-mode system.  Here we discuss how to derive the parameters in $\hat{H}$ in an optical lattice, where the initial state of Bosons occupies the zero momentum states of two bands in an 1D optical lattice. The Hamiltonian in a 3D optical lattice reads
\begin{eqnarray}
\begin{split}
H & =  \int d {\bf r} \Psi^{\dag}({\bf r})\left(-\frac{\hbar^{2}\nabla^{2}}{2m}+V_{\text{op}}({\bf r})\right)\Psi({\bf r})\\
&+\frac{2\pi \hbar^{2}a_{s}}{m}\int d{\bf r}\Psi^{\dag}({\bf r})\Psi^{\dag}({\bf r})\Psi({\bf r})\Psi({\bf r}),\label{3D-Hamiltonian}
\end{split}
\end{eqnarray}
where $V_{\text{op}}({\bf r})=V_{\|}\sin^{2}(\frac{\pi x}{d})+V_{\bot}\left(\sin^{2}(\frac{\pi y}{d})+\sin^{2}(\frac{\pi z}{d})\right)$. For large enough $V_{\bot}$,  the system is divided into independent 1D tubes. The wavefunction in the $y-z$ plane is a $s$-band Wannier function. In the $x$ direction, we consider the $s$ and $d$ bands, as realized in experiment \cite{zhou-13, zhou-16},  that shows occupation in the $p$ band is negligible in relevant experimental time scales. Our results can be straightforwardly generalized to bosons occupying the $s$ and $p$ bands. 
$\Psi({\bf r})$ reduces to $
\Psi({\bf r}) = \sum_{i,\sigma=s,d }\psi_{\sigma}({\bf r}-{\bf R}_{i})b_{\sigma, i}=\sum_{i,\sigma=s,d}\left(\prod_{r=y,z}w_{\sigma}(x-x_{i}, V_{\|})w_{s}(r-r_{i}, V_{\bot})\right) b_{\sigma, i}$
and the Hamiltonian is rewritten as $ H=\sum_{\sigma=s,d}H_{\sigma}+H_{sd}+E_g N_{d}$  with $E_g$ the band gap, $H_{\sigma}$ the single band Bose-Hubbard model and $H_{sd}$ the coupling between the two bands,  i.e.,
\begin{eqnarray*}
\begin{split}
H_{\sigma}&=t_{\sigma}\sum_{i}(b^{\dag}_{\sigma,i}b_{\sigma,i+1}+h.c) +\frac{U_{\sigma}}{2}\sum_{i}n_{\sigma,i}(n_{\sigma,i}-1),\\
H_{sd} &=4U_{sd}\sum_{i}n_{s,i}n_{d,i}+U_{sd}\sum_{i}\left(b^{\dag}_{s,i}b^{\dag}_{s,i}b_{d,i}b_{d,i}+h.c \right)\\
&+V_{a}\sum_{i}\left(b^{\dag}_{s,i}b^{\dag}_{s,i}b_{s,i}b_{d,i}+h.c \right)\\
&+V_{b}\sum_{i}\left(b^{\dag}_{d,i}b^{\dag}_{d,i}b_{d,i}b_{s,i}+h.c \right)
\end{split}
\end{eqnarray*}
where $t_{\sigma} = \int d{\bf r}\psi^{\ast}_{\sigma}({\bf r}-{\bf R}_{i})\left(-\frac{\hbar^{2}\nabla^{2}}{2m}+V_{\text{op}}({\bf r})\right) \psi_{\sigma}({\bf r}-{\bf R}_{i+1})$ is the tunneling and  $U_{\sigma}=\frac{4\pi \hbar^{2}a_{s}}{m}\int d{\bf r} |\psi_{\sigma}({\bf r})|^{4}$ is the on-site interaction strength. $U_{sd}=\frac{2\pi \hbar^{2}a_{s}}{m}\int d {\bf r}|\psi_{s}({\bf r})|^{2}|\psi_{d}({\bf r})|^{2}$ and  $V_{a(b)}=\frac{4 \pi \hbar^{2}a_{s}}{m}\int d{\bf r} \psi^{3}_{s(d)}({\bf r})\psi_{d(s)}({\bf r}), 
$ are the coupling strength between the $s$ band and $d$ band. Fig.(\ref{s1}) shows realistic parameters for Rb in a 3D lattice.  

As the initial state occupies only zero crystal momentum, significant depletions to finite momentum states only emerge in a much larger time scale that is not relevant to our discussions, as the intra-band scattering to a finite momentum $\sim N$ is much weaker than the inter-band interactions at zero momentum $\sim N^2$. An effective Hamiltonian after projecting $\hat{H}$ to the zero crystal momentum states can be obtained,  as shown in Eq.(2) of the main text, where $g_{\sigma}=\frac{U_{\sigma}}{2L_\|}, g_{sd}=\frac{U_{sd}}{L_\|}, u=\frac{V_a}{L_\|}, u'=\frac{V_b}{L_\|}$, $L_\|$ is the number of lattice site in the $x$ direction, $N=\frac{N_{\text{tot}}}{L_{\bot}^2}$ is the number of bosons in a single 1D tube, and $L_{\bot}^2$  the total site number in the $y-z$ plane.

\section{$m$-particle correlation function}
\label{sec:B} 
\begin{figure*}[t]
\begin{center}
\includegraphics[width=7in]{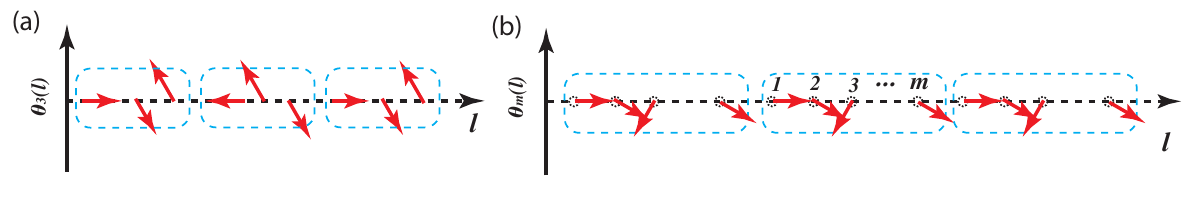}
\end{center}
\caption{(a) The direction of each arrow represent the phase $\theta_3(l)$ of a Fock state at $t=T_3$. Three concessive Fock states form a domain. (b) $m$ Fock states form a domain. In each domain,  destructive interference leads to vanishing $n$-body correlations if $n<m$.}
\label{s3}
\end{figure*}
Within the zero tunneling approximation, analytic results  of multi-particle correlation functions are available. For instance,  three-body correlation function $g_3(t)$ has the following analytic form
\begin{eqnarray}
\begin{split}
g_{3}(t)&=\frac{N(N-1)(N-2)}{8}\\
&\times \sum^{\infty}_{n=-\infty} e^{-\frac{N-3}{2}(\frac{3A_{2}t}{\hbar}-n\pi)^{2}}e^{i3 n\pi}e^{i\frac{3A_{1}t}{\hbar}},
\label{g3A}
\end{split}
\end{eqnarray}
where the decoherence time and revival time are
\begin{eqnarray}
\tau_3=\frac{\hbar}{3\sqrt{N-3}A_2}, \,\,\, T_3=\frac{\pi \hbar}{3A_2}.
\end{eqnarray}
At $t=T_3=\frac{\pi \hbar}{3A_2}$, 
\begin{eqnarray}
|\Psi(t=\frac{\pi \hbar}{3A_2})\rangle=\sum^{\frac{N}{2}}_{l=-\frac{N}{2}}\left(\frac{C_N^{\frac{N}{2}-l}}{2^N}\right)^{\frac{1}{2}}e^{-i\frac{A_1}{3A_2}\pi l}\theta_3(l),
\end{eqnarray}
where $\theta_3(l)=e^{-i\frac{\pi l^2}{3}}$.  $e^{-i\frac{A_1}{3A_2}\pi l}$ does not affect the amplitude of correlation functions, and  $\theta_3(l)$ is crucial. 

Consider one-particle correlation function,   $g_1(t)=\sum_{l}\alpha_{l}(t)\alpha^{\ast}_{l+1}(t)\mathcal{W}_{l}$, 
the behaviour of $g_1(t)$ can be understood from $\sum_{l}\theta^{\ast}_3(l+1)\theta_3(l)$ for the amplitudes of $\alpha_{l+p}(t)$ and $\alpha_l(t)$ are almost equal if  $p$  is not large, so the absolute value of $\alpha_l(t)\alpha^{\ast}_{l+1}(t) {\cal{W}}_{l}$ can be considered as a constant.
 At $t=\frac{\pi \hbar}{3A_2}$,  it is straightforward to show that $\sum^{l_0+2}_{l=l_0}\theta^{\ast}_3(l+1)\theta_3(l)=0$ for any $l_0$. If the 1D system is divided into domains, each of which contains three sites,  as shown in Fig.\ref{s3}(a), the destructive interference in each domain leads to vanishing  $g_1(t)$ when $t=T_3$. Similarly,  {$g_2(t=\frac{\pi \hbar}{3A_2})\approx \sum_{n} \sum^{l_0+2}_{l=l_0}\theta^{\ast}_3(l+2)\theta_3(l)$ where $n$ is the domain number. It also vanishes at $t=T_3$.  As for $g_3(t)$, we obtain  $\sum^{l_0+2}_{l=l_0}\theta^\ast_3(l+3)\theta_3(l)=-3$, and the constructive interference leads to the maximized $g_3(t)$ at $t=T_3$. 

The above results can be straightforwardly generalized to $m$-particle correlations, $g_m(t)=\langle \Psi(t)|b^{\dag m}_s b^m_d |\Psi(t)\rangle$. At $t=T_m= \frac{\pi \hbar}{mA_2}$,  we have
\begin{eqnarray}
|\Psi(t=\frac{\pi \hbar}{mA_2})\rangle=\sum^{\frac{N}{2}}_{l=-\frac{N}{2}}\left(\frac{C_N^{\frac{N}{2}-l}}{2^N}\right)^{\frac{1}{2}}e^{-i\frac{A_1}{mA_2}\pi l}\theta_m(l),
\end{eqnarray}
where $\theta_{m}(l)=e^{-i\frac{\pi  l^2}{m}}$ as shown in Fig.\ref{s3}(b).  It is straightforward to show that $\sum^{l_0+m-1}_{l=l_0}\theta^{\ast}_m(l+n)\theta_m(l)=0$, if  $n<m$,  and $\sum^{l_0+m-1}_{l=l_0}\theta^{\ast}_m(l+n)\theta_m(l)=(-1)^{m}m$  if $n=m$. Thus, at $t=T_m$, all correlation functions $g_{n<m}(t=T_m)$ vanish and $g_m(t=T_m)$ reaches its maximum. Similar to $g_1(t)$ and $g_2(t)$ discussed in the main text,  we have numerically verified all above results concerning $g_{m}(t)$ by solving the full Hamiltonian including tunnelings. At short time scales up to a few $T_m$, the zero tunneling approximation well reproduce the exact results.

\section{finite tunneling}
\label{sec:C}
As the correlation functions are mainly determined by the wavefunction near $l=0$, due to the bosonic enhancement factors, the Hamiltonian including tunnelings can be well approximated by 
\begin{eqnarray}
\begin{split}
H & =  \mathcal{J}\sum_{l}(|l+1\rangle \langle l|+h.c)\\
&+\sum_{l}(A_{2}l^{2}+A_{1}l) |l\rangle \langle l|, 
\label{a}
\end{split}
\end{eqnarray}
where $\mathcal{J}=uN^2 /4$ and $A_{1}=-E_g$. This is essentially a ``flat-band" approximation that replaces the $l$-dependent tunneling by its value at $l=0$. We have used the fact that $u$ is much larger than $u'$.  Hamiltonian (\ref{a}) can be diagonalized and $\psi_{m}(l)$ and $E_{m}$  are the eigenstate and eigenvalue, respectively. When $A_2$ vanishes,  Hamiltonian.(\ref{a}) reduces to a Wannier ladder, the eigenstates are Bessel functions and the energy spectrum is linear.   As both $|A_{1}|$ and $\mathcal{J}$ are much larger than $A_2$,  $A_2$ can be considered as a perturbation. The zero order wave function is $\psi^{(0)}_{m}(l)=J_{m+l}(x)$ with $x=2\mathcal{J}/E_g$ and  $E^{(1)}_{m}=-A_{1}m+A_{2}m^2$ if we consider the first order correction to the energy. Using (\ref{a}) and the initial state described in the main text,  the time-dependent state is
\begin{eqnarray}
|\Psi(t)\rangle & = & \sum_{m}\beta_{m}(t)|m\rangle =\sum_{l}\alpha_{l}(t)| l \rangle,
\label{b1}
\end{eqnarray}
where  $\alpha_{l}(t)  =  \sum_{m}\beta_{m}(t)\psi_{m}(l)$  and $\beta_{m}(t)=\beta_{m}(0)e^{-\frac{i}{\hbar} E_{m}t}$ with  $\beta_m(0)$ the overlap between the initial state and the eigenstate.
\subsection{One-particle correlation function}
Using Eq.(\ref{b1}), we obtain
\begin{eqnarray}
g_{1}(t) &=&\sum_{l}\sum_{mn}\beta^{\ast}_{m}(0)\beta_{n}(0)\psi^{\ast}_{m}(l+1)\psi_{n}(l)e^{\frac{i}{\hbar} (E_{m}-E_{n}) t} \mathcal{W}_l \nonumber\\
&=& g^{a}_1(t)+\mathcal{O}(A_2),
\label{f2}
\end{eqnarray}
where $\mathcal{W}_l=\sqrt{(\frac{N}{2}+l+1)(\frac{N}{2}-l)}$ and $g^{a}_1(t)$ is the zero order of $g_1(t)$, i.e.,
\begin{eqnarray}
\begin{split}
g^a_1(t) & = \sum_{l}\sum_{mn}\beta^{(0)}_{m}(0)\beta^{(0)}_{n}(0)\psi^{(0)}_{m}(l+1)\psi^{(0)}_{n}(l)\\
&\times e^{\frac{i}{\hbar} (E_{m}-E_{n})t} \mathcal{W}_l,
\label{eq2}
\end{split}
\end{eqnarray} 
where $\beta^{(0)}_m(0)=\langle \Psi(0)| \psi^{(0)}_m(l)\rangle$.   $\mathcal{W}_l$ can be approximated by $N/2$  and consider the  orthogonal property of Bessel functions,  i.e., $\sum^{\infty}_{n=-\infty}J_{n}(x)J_{n+q}(x)  =  \delta_{q,0}$,  we obtain

\begin{eqnarray}
\begin{split}
g^a_1(t)
&\approx \frac{N}{2}\sum_{m}\beta^{(0)}_{m}(0)\beta^{(0)}_{m+1}(0)e^{\frac{i}{\hbar} (E_{m}-E_{m+1})t}.
\label{ga}
\end{split}
\end{eqnarray}
In Eq.(\ref{ga}), $\beta^{(0)}_m(0)$ can be well approximated by
$\beta^{(0)}_{m}(0)\approx\left(\frac{2}{\pi N}\right)^{1/4}e^{-\frac{1}{N}(m-x)^{2}}$ with $x=2\mathcal{J}/E_g$. Substituting this equation into Eq.(\ref{ga}), replacing $E_{m}$ by $E^{(1)}_m$,  and using Poisson summation formula,  we obtain
\begin{eqnarray}
\begin{split}
g^{a}_1(t) & \approx  \frac{N}{2} e^{i \frac{A_{1}t}{\hbar}}e^{-\frac{1}{2N}}\\
&\times \sum_{n}e^{-\frac{N}{2}(n\pi -\frac{A_{2}t}{\hbar})^{2}}e^{i n\pi}e^{i(n\pi-\frac{A_{2}t}{\hbar})2x}.
\label{Ga1}
\end{split}
\end{eqnarray}
From the above equation, we see that the decoherence time $\tau_1=\hbar/(\sqrt{N}A_2)$ and the revival time $T_1=\hbar\pi/A_2$, consistent with the results in main text in large $N$ limit. Fig.\ref{s2}(a) shows the comparisons between Eq.(\ref{Ga1}) and the exact results,  which agree well at short times. 
\begin{figure*}
\begin{center}
\includegraphics[width=6.5in]{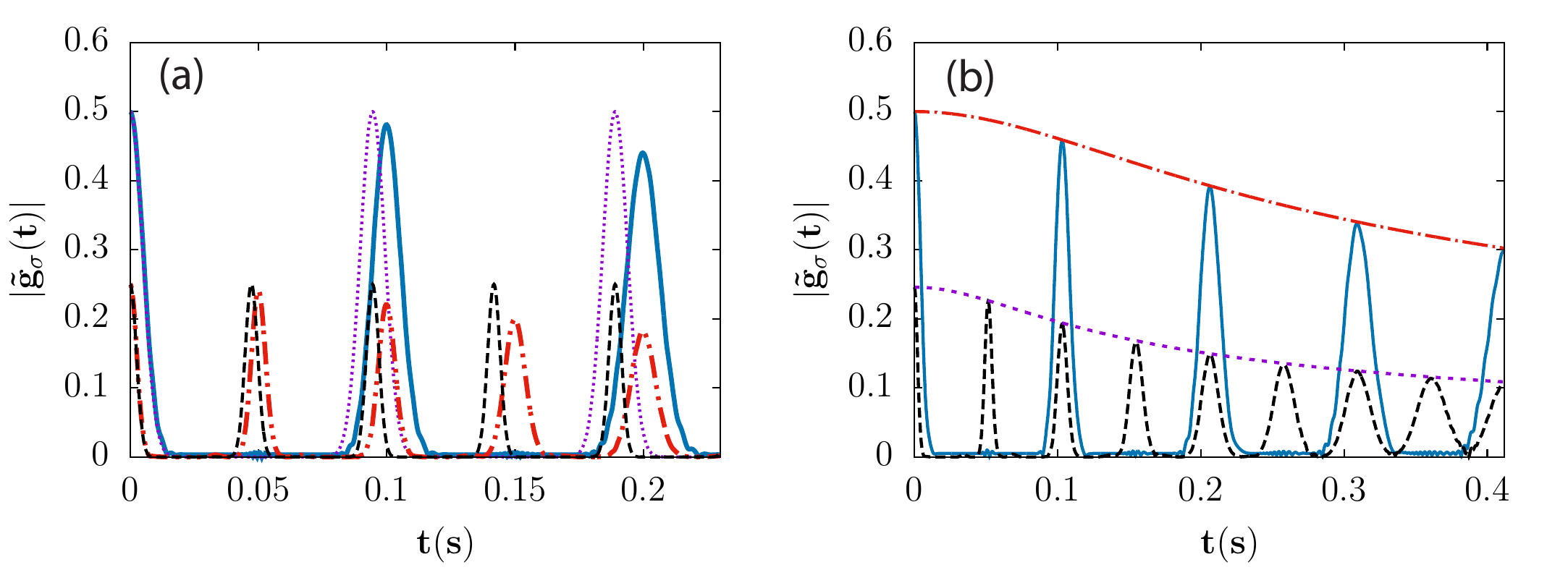}
\end{center}
\caption{(a) {One-particle and two-particle  correlation functions. Here  solid bule (dotted purple) and dashed-double-dotted  (dashed black) curves denote exact (analytic)  one-particle and two-particle correlation functions, respectively.
Parameters are  $N_{\text{tot}}=10^5, L_{\|}=L_{\bot}=50, N=40$. Here,  $\tilde{g}_{1}(t)=g_{1}(t)/N$ and $\tilde{g}_2(t)=g_2(t)/N^2$. (b) Envelops of correlation functions.  Solid blue (dashed black) curve represents exact one(two)-particle correlation function, the dashed dotted red(dotted purple) curve represents the envelop of one(two)-particle correlation functions. Parameters are $N_{\text{tot}}=1.5\times 10^5,  L_{\|}=L_{\bot}=50, N=60$,  $A_{1}=-7.6E_{R}$, here $A_2$ and $A_3$ are determined by fitting, $A_{2}=1.5\times 10^{-3} E_{R}, A_3=3.5\times 10^{-6}E_{R}$, $a_{s}=5.1nm, d=426nm$, $V_{\|}=10E_{R}$, and $V_{\bot}=20E_{R}$. $N=N_{\text{tot}}/(L_{\bot})^2$ is the particle number per 1D tube.}}
\label{s2}
\end{figure*}
In the zero tunneling approximation, we have seen that $g_1(t=T_2)=0$. In the full numerical calculations, there is a small  residue as the blue curve in Fig.\ref{s2}(a) shows. This small residue comes from high order correction from the tunneling. For instance, consider the following term $g^b_1(t)$,  
\begin{widetext} 
\begin{eqnarray}
\begin{split}
g^b_1(t) & =  \sum_{l}\sum_{mn}\beta^{(0)}_{m}(0)\beta^{(0)}_{n}(0)\left(\psi^{(0)}_{m}(l+1)\psi^{(1)}_{n}(l)+\psi^{(1)}_{m}(l+1)\psi^{(0)}_{n}(l)\right) e^{\frac{i}{\hbar} (E_{m}-E_{n})t} \mathcal{W}_l \\
&\approx \frac{N}{2}\sum_{l}\sum_{mn}\beta^{(0)}_{m}(0)\beta^{(0)}_{n}(0)
 \left(\psi^{(0)}_{m}(l+1)\psi^{(1)}_{n}(l)+\psi^{(1)}_{m}(l+1)\psi^{(0)}_{n}(l)\right)
e^{\frac{i}{\hbar} (E_{m}-E_{n})t},
\label{gb}
\end{split}
\end{eqnarray}
\end{widetext}
where $\psi^{(1)}_{m}(l)$  is the first order correction of the eigenstate and $\mathcal{W}_l$ have been replaced by $N/2$. Due to orthogonality of Bessel function,  Eq.(\ref{gb}) reduces to the following expression, 
\begin{eqnarray}
\begin{split}
g^{b}_1(t) &=\frac{N}{2}\frac{A_2}{Eg}\sum_{m n}\beta^{(0)}_m(0)  \beta^{(0)}_n(0)\\
&\times \left(\frac{\gamma_{m+1,n}}{n-(m+1)}+\frac{\gamma_{n-1,m}}{m-(n-1)}\right)e^{\frac{i}{\hbar}(E_m-E_n)t},
\label{gb1}
\end{split}
\end{eqnarray}
where $\gamma_{nm}=\sum_l l^2 J_{n+l}(x)J_{m+l}(x)$. Consider the leading term with $m=n$,  Eq.(\ref{gb1})  reduces to 
\begin{eqnarray}
g^{b}_1(t)\approx \frac{N}{2}\frac{A_2}{E_g}\sum_{m}|\beta^{(0)}_m(0)|^2 (-\gamma_{m+1,m}+\gamma_{m-1,m}),
\end{eqnarray}
where $\gamma_{m,m+1} =  -\frac{1}{2}x(1+2m)$ and $\gamma_{m-1,m}  =-\frac{1}{2}x\left(1+2(m-1)\right)$. Thus $g^b_1(t) \approx  N \frac{\mathcal{J} A_2 }{E^2_g} $. With decreasing tunneling down to zero, the residue vanishes.

\subsection{two-particle and three-particle correlation functions}
Using  Eq.(\ref{b1}),  we obtain
\begin{eqnarray}
g_2(t)  &=&\sum_{l}\sum_{mn}\beta^{\ast}_{m}(t)\beta_{n}(t)\psi^{\ast}_{m}(l+2)\psi_{n}(l)e^{\frac{i}{\hbar}(E_m-E_n)t}\mathcal{V}_{l} \nonumber\\
&=&g^{a}_2(t)+\mathcal{O}(A_2),
\label{G2}
\end{eqnarray}
where $\mathcal{V}_{l}=\sqrt{(\frac{N}{2}+l+2)(\frac{N}{2}+l+1)(\frac{N}{2}-l)(\frac{N}{2}-l-1)}$ 
and 
\begin{eqnarray}
 \begin{split}
g^a_2(t)&=\sum_{l}\sum_{mn}\beta^{(0)}_{m}(t)\beta^{(0)}_{n}(t)\psi^{(0) }_{m}(l+2)\psi^{(0)}_{n}(l)\\
&\times
e^{\frac{i}{\hbar}(E_m-E_n)t}\mathcal{V}_{l},\\
\end{split}
\label{g2A}
\end{eqnarray}
Using similar techniques in calculations of $g_1(t)$,   Eq.(\ref{g2A}) can be rewritten as
\begin{eqnarray}
\begin{split}
g^a_2(t)&=\frac{N^2}{4}e^{i\frac{2 A_1 t}{\hbar}}e^{-\frac{2}{N}}\\
&\times \sum_{n}e^{-\frac{N}{2}(n\pi-\frac{2A_2 t}{\hbar})^2}e^{i(n\pi-\frac{2A_2 t}{\hbar})2x}.
\label{g2A2}
\end{split}
\end{eqnarray}
Fig.\ref{s2}(a) shows the comparison  between analytic and exact results, which match well at small $t$.

The three-particle correlation function is written as,
\begin{eqnarray}
\begin{split}
g_3(t)&=\sum_{l}\sum_{mn}\beta^{\ast}_{m}(0)\beta_{n}(0)\psi^\ast_{m}(l+3)\psi_{n}(l)e^{i(E_m-E_n)t}\mathcal{U}_{l}\\
&=g^{a}_3(t)+\mathcal{O}(A_2).
\end{split}
\label{g3}
\end{eqnarray}
where
\begin{eqnarray}
\begin{split}
\mathcal{U}_{l}&=\sqrt{(\frac{N}{2}+l+3)(\frac{N}{2}+l+2)(\frac{N}{2}+l+1)}\\
&\times \sqrt{(\frac{N}{2}-l)(\frac{N}{2}-l-1)(\frac{N}{2}-l-2)}
\end{split}
\end{eqnarray}
 and 
\begin{eqnarray}
 \begin{split}
g^a_3(t)&=\sum_{l}\sum_{mn}\beta^{(0)}_{m}(t)\beta^{(0)}_{n}(t)
 \psi^{(0)}_{m}(l+3)\psi^{(0)}_{n}(l)\\
&\times e^{\frac{i}{\hbar}(E_m-E_n)t}
\mathcal{U}_{l}\\
&=\frac{N^{3}}{8}e^{i\frac{3A_1 t}{\hbar}}e^{-\frac{9}{2N}}\\
&\times \sum_n e^{-\frac{N}{2}(n\pi-\frac{3A_2 t}{\hbar})^2}e^{i n \pi}e^{i(n\pi-\frac{3A_2 t}{\hbar})2x}
\end{split}
\label{g21}
\end{eqnarray}
$\tau_{2,3}$ and $T_{2,3}$ derived from the above equations are consistent with those in the main text.

\section {Incomplete revival at long times}
\label{sec:D}
All previous analytical solutions show that $g_m(t)$ recovers its initial value at the revival time $t=T_m$. From the exact numerical calculations, we see that at short times, this is indeed true. However, at long times, the revival is not complete, i.e., the peaked value (or the envolop) of $g_m(t)$  gradually decreases as time goes on. This comes from high order corrections to the eigenenergies. The leading corrections is a cubic term  $\sim l^3$. For simplicity, we consider again the zero tunneling limit, and  the effective Hamiltonian reads
\begin{eqnarray}
H& = & \sum_{l}\Big(A_{3}l^{3}+A_{2}l^{2}+A_{1}l\Big)| l\rangle \langle l |, \label{A3-3D}
\end{eqnarray}
where $A_{1}=(g_s-g_d)N-E_g$, and $A_{2}$ and $A_{3}$ are determined by fitting the exact one-particle correlation function based on two consideration, one is  $A_2$ and $A_3$ are two small can not well determined by directly fitting the energy spectrum, the other is that within zero tunneling approximation, analytic results are available  as following shows, one can use the analytic results to fit the exact results and $A_2$ and $A_3$ can be well determined.

Using Hamiltonian (\ref{A3-3D}),  one-particle and two-particle correlation functions can be analytically obtained,
\begin{widetext}
\begin{eqnarray}
g_{1}(t)& =&  \frac{N}{2}\sqrt{\frac{2}{2-i\frac{3A_{3}t}{\hbar}(N-1)}}
\sum^{\infty}_{n=-\infty}e^{-\frac{N-1}{2-i\frac{3A_{3}t}{\hbar}(N-1)}(\frac{(3A_{3}+2A_{2})t}{2\hbar}+\frac{i}{N-1}-n \pi)^2} 
 e^{-\frac{1}{2(N-1)}}e^{i\frac{A_{1}+A_{2}+A_{3}}{\hbar}t }, \label{g1-A3}\\
g_{2}(t) & =&  \frac{N(N-1)}{4}\sqrt{\frac{1}{1-i\frac{3A_{3}t}{\hbar}(N-2)}}
 \sum^{\infty}_{n=-\infty}e^{-\frac{N-2}{2-i\frac{6A_{3}t}{\hbar}(N-2)}(\frac{(6A_{3}+2A_{2})t}{\hbar}+\frac{2i}{N-2}-n\pi)^2 }
 e^{-\frac{2}{N-2}}e^{i\frac{8A_{3}+4A_{2}+2A_{1}}{\hbar}t}.\label{g2-A3}
\end{eqnarray}
\end{widetext}
When $A_{3}$ is zero, Eq.(\ref{g1-A3}) and (\ref{g2-A3}) reduce to Eq.(\ref{g1}) and (\ref{g2}). For a finite $A_3$, envelops of correlation functions can be written as
\begin{eqnarray}
|g_{1}(t)| & \sim  & \frac{N}{2}\Big|\sqrt{\frac{2}{2-i \frac{3A_{3}t}{\hbar}(N-1)}} \Big| , \label{3D-envelop-1}\\
|g_{2}(t)| & \sim & \frac{N(N-1)}{4}  \Big|\sqrt{\frac{1}{1-i\frac{3A_{3}t}{\hbar}(N-2)}}\Big|. \label{3D-envelop-2}
\end{eqnarray}
Fig.\ref{s2}(b) shows that Eq.(\ref{3D-envelop-1}) and Eq.(\ref{3D-envelop-2}) well describe the envelopes of the exact results of $|g_{1}(t)|$ and $|g_{2}(t)|$.

Using Eq.(\ref{g1-A3}),   the  revival time and the decoherence time of $g_1(t)$ can be written as 
\begin{eqnarray}
\begin{split}
T_1  &=  \frac{\pi \hbar}{A_2+\frac{3}{2}A_3}, \,\,\,\,\,\,\,\, \\
 \tau^{n}_1 & =  \frac{\hbar }{\sqrt{N-1}A_2}\sqrt{1+(\frac{3A_3}{2A_2}n\pi (N-1))^2}, \label{tau-1}
 \end{split}
\end{eqnarray}
where the superscript $n$ denotes the value at the $n+1$th peak of $g_1(t)$. In the presence of small $A_3$, the width of the peaks increases with increasing $t$, i.e., decoherence time increases. 
Similar to $g_1(t)$,  the decoherence time and revival time of $g_2(t)$ are
\begin{eqnarray}
\begin{split}
T_2 & =  \frac{\pi \hbar}{2A_2+6A_3}, \,\,\,\\
\tau^n_2  &=  \frac{ \hbar}{\sqrt{N-2}2A_2}\sqrt{1+(\frac{3A_3}{2A_2}n\pi(N-2))^2 }.
\end{split}
\end{eqnarray}
From the above discussion, we can find that Hamiltonian (\ref{A3-3D}) can well describe the change of peak width and peak value of correlation functions.

\end{document}